\documentclass[doublecol]{epl2} 

\title{Effective three-body interactions in nuclei}
\shorttitle{Effective three-body interactions in nuclei} %Insert here a short version of the title if it exceeds 70 characters

\author{P.~Van~Isacker\inst{1} \and I.~Talmi\inst{2}}
\shortauthor{P.~Van~Isacker and I.~Talmi}

\institute{                    
\inst{1}
GANIL,
CEA/DSM--CNRS/IN2P3,
BP~55027, F-14076 Caen Cedex 5, France\\
\inst{2}
Weizmann Institute of Science, Rehovot, Israel}

\abstract{
It is shown that the three-body forces in the $1f_{7/2}$ shell,
for which recently evidence was found
on the basis of spectroscopic properties
of the Ca isotopes and $N=28$ isotones,
can be most naturally explained as an effective interaction
due to excluded higher-lying shells,
in particular the $2p_{3/2}$ orbit.}

\begin{document}
\maketitle

According to the shell model,
the ground state of $^{40}$Ca
with $Z=20$ protons and $N=20$ neutrons
has completely filled $1s$, $1p$, $1d$ and $2s$ proton and neutron orbits.
Valence neutrons from $N=21$ on
occupy the $1f_{7/2}$ orbit which is closed at the magic number $N=28$.
Similarly, protons added to the magic nucleus $^{48}$Ca
occupy the proton $1f_{7/2}$ orbit.
A natural description in the shell model 
of nuclei in this mass region is therefore obtained
by restricting neutrons and protons to the $1f_{7/2}$ orbit,
an approach which was extensively used
in the past~\cite{Lawson57,Talmi57,Ginocchio63,Cullen64}.
From the energy levels of nuclei
where these neutron and proton orbits are filled,
it is evident that there must be strong perturbations from higher configurations.
The features predicted for relative positions of levels
by using two-body effective interactions
between the protons and between the neutrons,
are only roughly obeyed.
This is not surprising due to the proximity of higher orbits.
In $^{41}$Ca, the $3/2^-$ level,
associated with the $2p_{3/2}$ orbit,
lies only about 2 MeV above the $7/2^-$ ground state.
Effects of configuration mixing
have been considered many years ago
for the calcium isotopes~\cite{Engeland66,Federman66}
and for the $N=28$ isotones~\cite{Auerbach67,Lips70}
with results that turned out to be in rather good agreement
with experiment.

Effective interactions are determined by the model spaces for which they are intended.
If pure $(1f_{7/2})^n$ configurations are adopted,
two-body interactions do not yield good agreement with experiment.
Hence, three-body effective interactions have been introduced~\cite{Eisenstein73,Quesne70}.
This approach may well reproduce level energies in a better way.
The pure $1f_{7/2}$ nuclear states, however,
will be only an approximation to the correct shell-model states.
Even if effective operators, other than the Hamiltonian may provide some help,
they cannot provide information on more complicated processes like beta decay.
Recently, a detailed discussion of the effects of three-body interactions
on various spectroscopic observables was presented by Zelevinsky~\cite{Zelevinsky09}.

Recently, three-body interactions were introduced,
in addition to two-body ones, for dealing with the proton and neutron $1f_{7/2}$ shells~\cite{Volya09}.
The author considers his results as a ``manifestation of three-body forces in $1f_{7/2}$-shell nuclei".
The good agreement he obtains could be evidence of such forces
if the $1f_{7/2}$ shell was a pure one.
Even within the shell model, whose wave functions are not the real ones,
the $1f_{7/2}$ shells are strongly perturbed.
Such three-body interactions play a role only
if the space used is a pure $1f_{7/2}$ shell.
It may happen that introducing four-body and higher interactions
will improve the fit to the data~\cite{Quesne70}
but this will imply a severe loss of predictive power.
It seems to us that it is more reasonable
to consider possible interactions with higher configurations.
Attempts in this direction were made in the past
and good agreement with experiment has been obtained
by using only two-body forces.
In fact, we aim to show that the effect of mixing with nearby configurations
leads to a three-body effective interaction rather similar to those of ref.~\cite{Volya09}.

The effect of perturbations of the pure $1f_{7/2}$ shell,
due to two-body effective interactions,
may assume the form of additional two-body interactions
as well as three-body ones. 
Such effects were considered in the past in atomic spectroscopy~\cite{Rajnak63,Racah67}.
The aim of this note is to derive, by adopting a simple approximation,
these additional two-body and three-body interactions
arising from second-order perturbations.
In pure $(7/2)^n$ configurations,
all eigenstates have definite seniority for any two-body interaction.
There is experimental evidence that the two lowest $J=4$ levels in $^{52}$Cr
are mixtures of states with seniorities $v=2$ and $v=4$.
This seniority mixing was obtained in ref.~\cite{Volya09}
by using a three-body interaction.
This mixing, however, was obtained in refs.~\cite{Auerbach67,Lips70}
due to mixing of these states with those obtained
by raising one $1f_{7/2}$ proton into the $2p_{3/2}$ orbit.

Consider first the $j^n$  configuration.
The two-body interaction $\hat V$
is taken to be a perturbation on the single-nucleon Hamiltonian.
The first-order contribution of $\hat V$
is its expectation value in the states considered.
In second order, the contribution to the energy of the state $aJ$ of the $j^n$ configuration
is given by
\begin{equation}
\sum_c
\frac{\langle aJ|\hat V|cJ\rangle\langle cJ|\hat V|aJ\rangle}
{E(aJ)-E(cJ)},
\label{e_pert2}
\end{equation}
where the summation is over all states $cJ$
which have non-vanishing matrix elements with the state $aJ$.
Such states are in configurations
which differ from that of $aJ$ in single-nucleon states of at most two nucleons.
These are the $j^{n-1}j'$ and  $j^{n-2}j'j''$ configurations.
It is well known that contributions of the interaction with the latter configurations
yield, in second-order perturbation, effective two-body interactions.
In the following, the contributions of configurations $cJ$
which differ from the $aJ$ configuration by the state of one nucleon will be considered.

The energies of all states of the $j^{n-1}j'$ configuration $cJ$
may differ by small amounts in the difference $E(aJ)-E(cJ)$.
In such cases it may be a good approximation
to replace this difference by a constant term,
independent of the quantum numbers
which characterize the state with the given $J$ in the excited configuration $c$.
It is then possible to carry out the summation over all states $cJ$.
In the following we make this approximation
and we also assume that the energy differences in (\ref{e_pert2})
are the same for all states in the $aJ$ configuration
for any number $n$ of (identical) nucleons.
This is in accordance with the strict rules of perturbation theory
where these differences are determined only by the single-nucleon Hamiltonian.
It is a very good approximation for electron states in atoms;
in nuclei, it is not expected to be a very good one.

Using second quantized operators,
explicit expressions may be obtained for the resulting two-body and three-body operators.
The part of the Hamiltonian
which connects the $j^n$ and $j^{n-1}j'$ configurations
is given by~\cite{Talmi93}
\begin{equation}
\sum_{JM}
\langle j^2J|\hat V|jj'J\rangle
A^\dag(jj'JM)
A(jjJM),
\end{equation}
where $A(jj'JM)=\left(A^\dag(jj'JM)\right)^\dag$ and
\begin{equation}
A^\dag(jj'JM)=
\frac{1}{\sqrt{1+\delta_{jj'}}}
\sum_{mm'}(jmj'm'|jj'JM)a^\dag_{jm}a^\dag_{j'm'}.
\end{equation}
From the formula~(\ref{e_pert2}),
applied to the intermediate configurations $cJ$ of the type $jj'J$
and assuming equal denominators,
the following sums are obtained:
\begin{eqnarray}
&&\sum_{M_1M_2}
\langle j^2J_1|\hat V|jj'J_1\rangle
A^\dag(jjJ_1M_1)A(jj'J_1M_1)
\nonumber\\&&\quad\times
\langle j^2J_2|\hat V|jj'J_2\rangle
A^\dag(jj'J_2M_2)A(jjJ_2M_2).
\end{eqnarray}
With the help of the anti-commutation relations
between creation and annihilation operators
as well as some tensor algebra identities,
this expression can be evaluated.
The result is the sum of a two-body interaction
\begin{equation}
\sum_{JM}
|\langle j^2J|\hat V|jj'J\rangle|^2
A^\dag(jjJM)A(jjJM),
\end{equation}
and a three-body one
\begin{eqnarray}
&&\sum_{J_1J_2JM}
\langle j^2J_1|\hat V|jj'J_1\rangle
\langle j^2J_2|\hat V|jj'J_2\rangle
\Bigl\{\begin{array}{ccc}
j&J_1&j'\\j&J_2&J
\end{array}\Bigr\}
\nonumber\\&&\quad\times
[A^\dag(jjJ_1)\times a_j^\dag]^{(JM)}
[A(jjJ_2)\times a_j]^{(JM)}.
\end{eqnarray}

Instead of using these expressions,
it is simpler to reduce the matrix elements in the $j^n$ configuration
to those in the $j^2$ and $j^3$ configurations,
where the calculation of matrix elements is straightforward.
These matrix elements define uniquely the operators.
To calculate matrix elements
between states $aJ$ and $bJ$ in the $j^n$ configuration
one may use the expansion~\cite{Talmi93}
\begin{equation}
\Psi(j^naJ)=
\sum_{a_0J_0}
[j^{n-1}(a_0J_0)jJ|\}j^naJ]
\Psi(j^{n-1}(a_0J_0)j_nJ),
\end{equation}
where $[j^{n-1}(a_0J_0)jJ|\}j^naJ]$
are coefficients of fractional parentage (c.f.p.)
which express the anti-symmetric $n$-particle wave function $\Psi(j^naJ)$
in terms of an anti-symmetric $(n-1)$-particle wave function $\Psi(j^{n-1}a_0J_0)$
coupled with that of the $n^{\rm th}$ particle $j_n$
to total angular momentum $J$.
Since this wave function, as well as the one of the $bJ$ state, is fully antisymmetric,
it is possible to replace the matrix element of any two-body interaction $\hat V=\sum_{ik}\hat V_{ik}$
by $\hat V_{12}$  multiplied by $n(n-1)/2$, the number of such terms.
Thus, we obtain
\begin{eqnarray}
\lefteqn{\langle j^naJ|\hat V|j^nbJ\rangle}
\nonumber\\&=&
\frac{n(n-1)}{2}
\langle j^naJ|\hat V_{12}|j^nbJ\rangle
\nonumber\\&=&
\frac{n(n-1)}{2}
\sum_{a_0b_0J_0}
\langle j^{n-1}a_0J_0|\hat V_{12}|j^{n-1}b_0J_0\rangle
\nonumber\\&&\times
[j^{n-1}(a_0J_0)jJ|\}j^naJ]
[j^{n-1}(b_0J_0)jJ|\}j^nbJ]
\nonumber\\&=&
\frac{n}{n-2}
\sum_{a_0b_0J_0}
\langle j^{n-1}a_0J_0|\hat V|j^{n-1}b_0J_0\rangle
\nonumber\\&&\times
[j^{n-1}(a_0J_0)jJ|\}j^naJ]
[j^{n-1}(b_0J_0)jJ|\}j^nbJ].
\end{eqnarray}
This procedure may be further applied
to the matrix elements in the $j^{n-1}$ configuration
until the $j^2$ configuration is reached. 

Similarly, it is possible to apply this procedure
to any three-body interaction $\hat V=\sum_{ijk}\hat V_{ijk}$
and to replace it by $n(n-1)(n-2)\hat V_{123}/6$.
Using c.f.p., it is possible  to calculate matrix elements in the $j^n$ configuration
in terms of matrix elements in the $j^{n-1}$ configuration.
Thus, we obtain
\begin{eqnarray}
\lefteqn{\langle j^naJ|\hat V|j^nbJ\rangle}
\nonumber\\&=&
\frac{n(n-1)(n-2)}{6}
\langle j^naJ|\hat V_{123}| j^nbJ\rangle
\nonumber\\&=&
\frac{n(n-1)(n-2)}{6}
\sum_{a_0b_0J_0}
\langle j^{n-1}a_0J_0|\hat V_{123}|j^{n-1}b_0J_0\rangle
\nonumber\\&&\times
[j^{n-1}(a_0J_0)jJ|\}j^naJ]
[j^{n-1}(b_0J_0)jJ|\}j^nbJ]
\nonumber\\&=&
\frac{n}{n-3}
\sum_{a_0b_0J_0}
\langle j^{n-1}a_0J_0|\hat V|j^{n-1}b_0J_0\rangle
\nonumber\\&&\times
[j^{n-1}(a_0J_0)jJ|\}j^naJ]
[j^{n-1}(b_0J_0)jJ|\}j^nbJ].
\end{eqnarray}
This procedure may be continued until the $j^3$ configuration is reached.

Let us now turn to the calculation of the effective interaction
in the $j^2$ and $j^3$ configurations,
due to additional shells $j'$
that are not explicitly taken into account.
The two-body operators may be obtained
from perturbation theory in the two-nucleon configuration.
The contribution in second-order perturbation
to the $j^2$ configuration in a state with spin $J_0$
can be simply calculated from (\ref{e_pert2})
and is given by
\begin{equation}
\sum_{j'}\frac{|\langle j^2J_0|\hat V|jj'J_0\rangle|^2}
{E(j^2J_0)-E(jj'J_0)},
\end{equation}
where the matrix elements of the interaction $\hat V$ 
are between anti-symmetric and normalized states
$\langle j^2J_0|$ and $|jj'J_0\rangle$.

To obtain the matrix elements of the three-nucleon contribution,
the $j^3$ configuration should be considered.
Results of perturbations of states in the $j^3$ configuration
may be calculated as follows.
Anti-symmetric and normalized $cJ$ states may be expressed as
\begin{eqnarray}
\Psi(j^2(J_1)j'J)&=&
\frac{1}{\sqrt3}
\bigl(\Psi(j^2(J_1)j'_3 J)-\Psi(j^2(J_1)j'_1 J)
\nonumber\\&&\qquad
-\Psi(j^2(J_1)j'_2 J)\bigr).
\end{eqnarray}
The matrix element of  $\hat V=\hat V_{12}+\hat V_{13}+\hat V_{23}$
between the fully anti-symmetric states
$|aJ\rangle\equiv|j^3aJ\rangle$
and $|cJ\rangle\equiv|j^2(J_1)j'J\rangle$
is equal to the matrix element of $3\hat V_{12}$ and hence is given by
\begin{eqnarray}&&
-\sqrt{3}\sum_{J_0}
[j^2(J_0)jJ|\}j^3aJ]
\langle j^2(J_0)j_3 J|\hat V_{12}|j^2(J_1)j'_1J\rangle
\nonumber\\&&
-\sqrt{3}\sum_{J_0}
[j^2(J_0)jJ|\}j^3aJ]
\langle j^2(J_0)j_3J|\hat V_{12}|j^2(J_1)j'_2J\rangle.
\nonumber\\
\label{e_matint}
\end{eqnarray}
To evaluate the first sum it is convenient to carry out
a change of coupling transformation 
on the state $|j_3j_2(J_1)j'_1J\rangle=-|j_2j_3(J_1)j'_1J\rangle$,
\begin{eqnarray}
|j_2j_3(J_1)j'_1J\rangle&=&
\sum_{J_2}
(-1)^{j+j'+J_1+J_2}
%\sqrt{(2J_1+1)(2J_2+1)}
\hat J_1\hat J_2
\nonumber\\&\times&
\Bigl\{\begin{array}{ccc}
j&J_2&J\\j'&J_1&j
\end{array}\Bigr\}
|j_2j'_1(J_2) j_3J\rangle,
\end{eqnarray}
where $\hat J_i\equiv\sqrt{2J_i+1}$.
The integration over $j_3$ can be carried out
and yields the non-vanishing terms in which the equality $J_2=J_0$ must hold,
as follows
\begin{eqnarray}&&
\sqrt{3}(-1)^{j+j'}
\sum_{J_0}
%\sqrt{(2J_0+1)(2J_1+1)}
\hat J_0\hat J_1
[j^2(J_0)jJ|\}j^3aJ]
\nonumber\\&&\qquad\qquad\times
\Bigl\{\begin{array}{ccc}
j&J_0&J\\j'&J_1&j
\end{array}\Bigr\}
\langle j^2J_0|\hat V|j_2j'_1J_0\rangle.
\end{eqnarray}
Similarly, the second summation in~(\ref{e_matint}) yields the following result:
\begin{eqnarray}&&
-\sqrt{3}(-1)^{j+j'}
\sum_{J_0}
%\sqrt{(2J_0+1)(2J_1+1)}
\hat J_0\hat J_1
[j^2(J_0)jJ|\}j^3aJ]
\nonumber\\&&\qquad\qquad\times
\Bigl\{\begin{array}{ccc}
j&J_0&J\\j'&J_1&j
\end{array}\Bigr\}
\langle j^2J_0|\hat V|j_1j'_2J_0\rangle.
\end{eqnarray}
The two expressions may be combined yielding the result
\begin{eqnarray}
&&-\sqrt{6}(-1)^{j+j'}
\sum_{J_0}
%\sqrt{(2J_0+1)(2J_1+1)}
\hat J_0\hat J_1
[j^2(J_0)jJ|\}j^3aJ]
\nonumber\\&&\qquad\qquad\times
\Bigl\{\begin{array}{ccc}
j&J_0&J\\j'&J_1&j
\end{array}\Bigr\}
\langle j^2J_0|\hat V|jj'J_0\rangle,
\end{eqnarray}
where again the states $\langle j^2J_0|$ and $|jj'J_0\rangle$
are anti-symmetric and normalized.
To obtain the second-order contribution~(\ref{e_pert2}) of these perturbations,
this expression should be multiplied by a similar one
where the summation is over all values of $J'_0$.
The second-order perturbation contribution
in the $j^3$ configuration is therefore
\begin{eqnarray}&&
6\sum_{j'J_1}(2J_1+1)
\sum_{J_0J'_0}
\hat J_0\hat J'_0
%\sqrt{(2J_0+1)(2J'_0+1)}
\nonumber\\&&\qquad\times
[j^2(J_0)jJ|\}j^3aJ]
\Bigl\{\begin{array}{ccc}
j&J_0&J\\j'&J_1&j
\end{array}\Bigr\}
\nonumber\\&&\qquad\times
[j^2(J'_0)jJ|\}j^3aJ]
\Bigl\{\begin{array}{ccc}
j&J'_0&J\\j'&J_1&j
\end{array}\Bigr\}
\nonumber\\&&\qquad\times
\frac{\langle j^2J_0|\hat V|jj'J_0\rangle\langle jj'J'_0|\hat V|j^2J'_0\rangle}
{E(j^3aJ)-E(j^2(J_1)j'J)}.
\end{eqnarray}

\begin{table*}
\centering
\caption{
The $T=3/2$ three-body interaction matrix elements
in the $1f_{7/2}$ shell obtained by Volya
and the effective three-body interaction matrix elements
derived with the formula~(\ref{e_gen3})
for various p3f7 interactions and for the gxpf1a interaction.
Energies are in keV.}
\label{t_gen3}
\smallskip
\begin{tabular}{rrrrcrrrcr}
\hline\hline
&\multicolumn{3}{c}{$Z=20$}&~~~&\multicolumn{3}{c}{$N=28$}&~~~&\\
\cline{2-4}\cline{6-8}
$J$&Volya~\cite{Volya09}&p3f7~\cite{Engeland66}&p3f7~\cite{Federman66}&
&Volya~\cite{Volya09}&p3f7~\cite{Auerbach67}&p3f7~\cite{Lips70}&&gxpf1a\\
\hline
3/2& $-559~(273)$&$-412$&$-173$&&$-128~(88)$&$-53$&$-115$&&  $-41$\\
5/2& $2~(185)$     &$-207$&$-104$&&$-18~(70)$  &$-54$&$-57$  &&  $-75$\\
7/2& $53~(70)$     &$138$ &$76$    &&$55~(28)$   &$42$ &$37$   &&  $45$\\
9/2& $272~(98)$   &$392$ &$157$  &&$122~(41)$ &$36$ &$110$ &&$105$\\
11/2&$51~(130)$  &$29$   &$46$    &&$102~(43)$ &$72$ &$6$     &&  $79$\\
15/2&$-24~(73)$   &$77$   &$28$    &&$-53~(29)$ &$5$   &$23$    &&  $-1$\\
\hline\hline
\end{tabular}
\end{table*}
As explained above,
we make the approximation that the energy denominators in (\ref{e_pert2})
are the same for all states of the $j^n$ and $j^{n-1}j'$ configurations.
This means that for the $j^3$ configurations we assume that
the energies $E(j^3aJ)\equiv E(j^3)$ and $E(j^2(J_1)j'J)\equiv E(j^2j')$,
are independent of $a$, $J_1$ and $J$,
and given be $3\epsilon_j$ and $2\epsilon_j+\epsilon_{j'}$, respectively,
where $\epsilon_j$ and $\epsilon_{j'}$ are single-particle energies.
In this case, the summation over $J_1$ can be carried out directly
due to the identities of Racah coefficients.
The sum over even values of $J_1$ can be expressed
as a linear combination of sums in which the summation is over all values of $J_1$,
yielding
\begin{eqnarray}
\lefteqn{
2\sum_{J_1\;{\rm even}}
(2J_1+1)
\Bigl\{\begin{array}{ccc}
j&J_0&J\\j'&J_1&j
\end{array}\Bigr\}
\Bigl\{\begin{array}{ccc}
j&J'_0&J\\j'&J_1&j
\end{array}\Bigr\}}
\nonumber\\&&=
\sum_{J_1}
(2J_1+1)
\Bigl\{\begin{array}{ccc}
j&J_0&J\\j'&J_1&j
\end{array}\Bigr\}
\Bigl\{\begin{array}{ccc}
j&J'_0&J\\j'&J_1&j
\end{array}\Bigr\}
\nonumber\\&&+
\sum_{J_1}
(-1)^{J_1}(2J_1+1)
\Bigl\{\begin{array}{ccc}
j&J_0&J\\j'&J_1&j
\end{array}\Bigr\}
\Bigl\{\begin{array}{ccc}
j&J'_0&J\\j'&J_1&j
\end{array}\Bigr\}
\nonumber\\&&=
\frac{\delta_{J_0J'_0}}{2J_0+1}+
\Bigl\{\begin{array}{ccc}
j&J&J_0\\j&j'&J'_0
\end{array}\Bigr\}.
\label{e_sumrac}
\end{eqnarray}
Substituting this result into the expression above we find
\begin{eqnarray}&&
3\sum_{j'}
\sum_{J_0J'_0}
\hat J_0\hat J'_0
%\sqrt{(2J_0+1)(2J'_0+1)}
[j^2(J_0)jJ|\}j^3aJ]
[j^2(J'_0)jJ|\}j^3aJ]
\nonumber\\&&\qquad\times
\left[
\frac{\delta_{J_0J'_0}}{2J_0+1}+
\Bigl\{\begin{array}{ccc}
j&J&J_0\\j&j'&J'_0
\end{array}\Bigr\}
\right]
\nonumber\\&&\qquad\times
\frac{\langle j^2J_0|\hat V|jj'J_0\rangle\langle jj'J'_0|\hat V|j^2J'_0\rangle}
{E(j^3)-E(j^2j')}.
\label{e_tot3}
\end{eqnarray}

To obtain the matrix elements of the genuine (but effective) three-body interaction,
the contribution of the two-body interaction should be subtracted.
The latter is given by
\begin{equation}
3[j^2(J_0)jJ|\}j^3aJ]^2
\sum_{j'}
\frac{|\langle j^2J_0|\hat V_{12}|jj'J_0\rangle|^2}
{E(j^2J_0)-E(jj'J_0)}.
\label{e_gen2}
\end{equation}
Again we assume that the energies
$E(j^2J_0)\equiv E(j^2)=2\epsilon_j$
and $E(jj'J_0)\equiv E(jj')=\epsilon_j+\epsilon_{j'}$,
are independent of  $J_0$.
In that case, subtracting this contribution from~(\ref{e_tot3}),
we find the genuine three-body matrix element to be equal to 
\begin{eqnarray}&\displaystyle
3\sum_{j'}
\sum_{J_0J'_0}
\hat J_0\hat J'_0
%\sqrt{(2J_0+1)(2J'_0+1)}
[j^2(J_0)jJ|\}j^3aJ]
[j^2(J'_0)jJ|\}j^3aJ]&
\nonumber\\&\displaystyle\times
\Bigl\{\begin{array}{ccc}
j&J&J_0\\j&j'&J'_0
\end{array}\Bigr\}
\frac{\langle j^2J_0|\hat V|jj'J_0\rangle\langle jj'J'_0|\hat V|j^2J'_0\rangle}
{E(j^3)-E(j^2j')}.&
\label{e_gen3}
\end{eqnarray}

It should be pointed out that the second-order perturbation contribution
is contained in both two-body and three-body interactions.
The contribution of second order in perturbation theory is always attractive (negative).
This need not be the case for the three-body interaction
which may lead to some strange results.
For example, if $j=7/2$ and $j'=1/2$,
there is no contribution of this perturbation to the $j^3$ state with $J=15/2$.
The highest value of $J_1$ is $J_1=6$ which cannot couple with $j'=1/2$ to yield $J=15/2$.
Still, there is a non-vanishing contribution to the matrix element
with $J=15/2$ and $J_0=6$ of the two-body interaction.
This unphysical contribution is exactly cancelled
by the corresponding contribution from the three-body interaction. 
Such cancellation must take place in all $7/2^n$ configurations. 

In the calculation of matrix elements in the $j^3$ configuration,
the origin of the unphysical terms is clear.
The sum of products of two $6j$-symbols in (\ref{e_sumrac}) is always $1/(2J_0+1)$
even if all $6j$-symbols with $J_1$ even vanish.
In such a case, the result is due to symbols with odd $J_1$ values.
These symbols contribute to a non-vanishing symbol in which $J_1$ no longer appears.
If in (\ref{e_sumrac}) $J_0\neq J'_0$,
there is no contribution of terms with odd $J_1$ values to the three-body interaction.
If, however, $J_0=J'_0$,
the odd $J_1$ terms contribute with opposite signs the same amount to the two-body interactions
and to the three-body ones. 
Thus, the two-body (\ref{e_gen2}) and three-body (\ref{e_gen3}) expressions calculated above,
which are genuine two-body and three-body interactions, may contain unphysical terms.
The contributions of two-body interactions
are usually absorbed into the  effective two-body interaction.
If the interest is in three-body interactions
whose contributions cannot be mimicked by two-body terms,
the expressions obtained above should be used. 

The above formalism can be applied to the $1f_{7/2}$ shell.
Several interactions are available
that include effects from the $2p_{3/2}$ shell.
Two of them were derived on the basis of spectroscopic properties
of the calcium isotopes~\cite{Engeland66,Federman66}
and two more from those of the $N=28$ isotones~\cite{Auerbach67,Lips70}.
In particular, the authors of these references give numerical values
for the matrix elements involving the $2p_{3/2}$ shell
which enter the expression~(\ref{e_gen3})
and for the difference in single-particle energies,
$\epsilon_{2p_{3/2}}-\epsilon_{1f_{7/2}}$.
We refer to these interactions as p3f7,
followed by the relevant reference.
For comparison, we also include results obtained with
a more recent interaction which considers the $2p1f$ shell
and results from an empirical fit
(starting from a microscopically derived set of matrix elements)
to reproduce a large body of energy data in the $2p1f$-shell nuclei.
This is the gxpf1a interaction~\cite{Honma04}
which adopts the single-particles energies
$\epsilon_{1f_{7/2}}=-8.6240$,
$\epsilon_{2p_{3/2}}=-5.6793$,
$\epsilon_{2p_{1/2}}=-4.1370$
and $\epsilon_{1f_{5/2}}=-1.3829$~MeV.

We have used these interactions, without any modification,
to calculate the effective three-body interaction
which is induced when the model space is restricted to $1f_{7/2}$.
The results are shown in Table~\ref{t_gen3}.
For $j=7/2$ a three-nucleon state
is completely specified by its total angular momentum $J$
and there is no need for the additional label $a$.
There are six three-nucleon states
with $J=3/2$, 5/2, 7/2, 9/2, 11/2 and 15/2,
and each of these states defines a component of the three-body interaction.
According to the expression~(\ref{e_gen3}),
the total effective three-body interaction
results from additive contributions of the different shells $j'$
which can be $2p_{1/2}$, $2p_{3/2}$ or $1f_{5/2}$.
The three-body matrix elements of Volya~\cite{Volya09}
result from separate fits to the calcium isotopes
and to the $N=28$ isotones,
and they should thus be compared with corresponding matrix elements
obtained with the p3f7 interactions from refs.~\cite{Engeland66,Federman66}
and \cite{Auerbach67,Lips70}, respectively.
The numbers in parentheses in Table~\ref{t_gen3}
are also taken from Volya
and correspond to the variances of the parameters in the fits.

There are substantial variations
in the calculated effective three-body matrix elements.
In particular, those derived from the $Z=20$ p3f7 interactions
are generally larger than those obtained with the $N=28$ p3f7 interactions.
This results from a combination
of a larger difference $\epsilon_{2p_{3/2}}-\epsilon_{1f_{7/2}}$
and smaller two-body matrix elements
$\langle(1f_{7/2})^2J_0|\hat V|1f_{7/2}2p_{3/2}J_0\rangle$
in the latter interactions.
Note also that the effective three-body matrix elements
derived from gxpf1a are closer to those
obtained with the $N=28$ p3f7 interactions.

There are certainly differences
between these calculations and the results of Volya.
However, in spite of these differences and uncertainties,
it is clear that the matrix elements are correlated:
with one exception,
attractive (repulsive) matrix elements in the analysis of Volya
turn out to be attractive (repulsive) in our analysis.
The exception concerns the $J=15/2$ interaction
which is attractive in Volya's analysis
while it is repulsive for the p3f7 interactions
and essentially zero in gxpf1a.

It is believed that there are important {\it ab initio} three-body interactions
and also three-body interactions due to short-range correlations between nucleons.
The latter arise from admixtures of highly excited configurations.
So far, no evidence was found for the effects of three-body interactions
on states of valence nucleons.
In cases where rather pure shell-model configurations were observed,
states and energies were well determined by effective two-body interactions.
Clearly, the $1f_{7/2}$ shell for protons and for neutrons is not in this category
since there are low-lying configurations
whose states may well mix with the $(1f_{7/2})^n$ states.
Insisting on using pure $1f_{7/2}$ configurations,
better agreement with experiment is obtained
by incorporating effective three-body interactions.
We do not question the existence of such three-body interactions
but we surmise that most likely they arise from renormalization effects
due to  admixtures of rather low-lying configurations.
Our conclusion is that claims of evidence
for the existence of genuine three-nucleon interaction
should be treated with some skepticism
if they are based on calculations in a shell-model space
which seems to be too restricted.

%\acknowledgments
%Insert here the text.


\begin{thebibliography}{00}

\bibitem{Lawson57}
\Name{Lawson R.D. \and Uretsky J.L.}
\REVIEW{Phys. Rev.}{106}{1957}{1369}

\bibitem{Talmi57}
\Name{Talmi I.}
\REVIEW{Phys. Rev.}{107}{1957}{326}

\bibitem{Ginocchio63}
\Name{Ginocchio J.N. \and French J.B.}
\REVIEW{Phys. Lett.}{7}{1963}{137}

\bibitem{Cullen64}
\Name{McCullen J.D., Bayman B.F. \and Zamick L.}
\REVIEW{Phys. Rev.}{134}{1964}{B515}

\bibitem{Engeland66}
\Name{Engeland T. \and Osnes E.}
\REVIEW{Phys. Lett.}{20}{1966}{424}

\bibitem{Federman66}
\Name{Federman P. \and Talmi I.}
\REVIEW{Phys. Lett.}{22}{1966}{469}

\bibitem{Auerbach67}
\Name{Auerbach N.}
\REVIEW{Phys. Lett. B}{24}{1967}{260}

\bibitem{Lips70}
\Name{Lips K. \and McEllistrem M.T.}
\REVIEW{Phys. Rev. C}{1}{1970}{1009}

\bibitem{Eisenstein73}
\Name{Eisenstein I. \and Kirson M.}
\REVIEW{Phys. Lett. B}{47}{1973}{315}

\bibitem{Quesne70}
\Name{Quesne C.}
\REVIEW{Phys. Lett. B}{31}{1970}{7}

\bibitem{Zelevinsky09}
\Name{Zelevisnky V.}
\REVIEW{Yadernaya Fizika [Phys. At. Nucl.]}{72}{2009}{1107}

\bibitem{Volya09}
\Name{Volya A.}
\REVIEW{Eur. Phys. Lett.}{86}{2009}{52001}

\bibitem{Rajnak63}
\Name{Rajnak K. \and Wybourne B.G.}
\REVIEW{Phys. Rev.}{132}{1963}{280}

\bibitem{Racah67}
\Name{Racah G. \and Stein J.}
\REVIEW{Phys. Rev.}{156}{1967}{58}

\bibitem{Talmi93}
\Name{Talmi I.}
\Book{Simple Models of Complex Nuclei.
The Shell Model and Interacting Boson Model}
\Publ{Harwood, Academic, Chur}
\Year{1993}

\bibitem{Honma04}
\Name{Honma M., Otsuka T., Brown B.A. \and Mizusaki T.}
\REVIEW{Phys.\ Rev.\ C}{69}{2004}{034335}

\end{thebibliography}
\end{document}